\documentclass[sigchi]{acmart}

\AtBeginDocument{%
  \providecommand\BibTeX{{%
    \normalfont B\kern-0.5em{\scshape i\kern-0.25em b}\kern-0.8em\TeX}}}

\setcopyright{rightsretained}

\copyrightyear{2019} 
\acmYear{2019} 
\acmConference[UbiComp/ISWC '19 Adjunct]{Adjunct Proceedings of the 2019 ACM International Joint Conference on Pervasive and Ubiquitous Computing and the 2019 International Symposium on Wearable Computers}{September 9--13, 2019}{London, United Kingdom}
\acmBooktitle{Adjunct Proceedings of the 2019 ACM International Joint Conference on Pervasive and Ubiquitous Computing and the 2019 International Symposium on Wearable Computers (UbiComp/ISWC '19 Adjunct), September 9--13, 2019, London, United Kingdom}\acmDOI{10.1145/3341162.3343783}
\acmISBN{978-1-4503-6869-8/19/09}

\begin{document}

\title{Poster: IoT Skullfort: Exploring the Impact of Internet Connected Cosplay}

\author{Rhys Beckett}
\affiliation{%
  \institution{Cardiff University}
  \city{Cardiff}
  \country{United Kingdom}}
\email{BeckettR@cardiff.ac.uk}

\author{Charith Perera}
\affiliation{%
  \institution{Cardiff University}
  \city{Cardiff}
  \country{United Kingdom}}
\email{charith.perera@acm.org}

\renewcommand{\shortauthors}{Beckett and Charith, et al.}

\begin{abstract}
In this paper, we explore the potential impact of Internet of Things (IoT) technology may have on the cosplay community. We developed a costume (an IoT Skullfort) and embedded IoT technology to enhance its capabilities and user interactions.  Sensing technologies are widely used in many different wearable domains including cosplay scenarios. However, in most of these scenarios, typical interaction pattern is that the costume responds to its environment or the player's behaviour (e.g.,  colour of  lights may get changed when  player moves hands). In contrast, our research focuses on exploring scenarios where the audience (third party) get to manipulate the costume behaviour (e.g., the audience get to change the colour of the Skullfort using a mobile application).  We believe such an audience (third party) influenced cosplay  brings new opportunities for enhanced entertainment. However, it also creates significant challenges. We report the results gathered through a focus group conducted in collaboration with cosplay community experts.
\end{abstract}

\ccsdesc[500]{Human-centered computing~Ubiquitous and mobile computing}

\keywords{Internt of Things, Cosplay, Fabricated Costumes}


\maketitle

\section{Introduction}

 Cosplay is the fusion of the words costume and play. It can describe the performance art in which individuals wear costumes to represent a specific character, or it can also refer to the costumes themselves. Those who participate in the art, are referred to as \textit{`Cosplayers'} \cite{Lamerichs2013}. Individuals can engage in cosplay in a variety of different ways. Most participants mainly engage by attending conventions whilst in costume. These conventions vary in size and theme; they can be smaller, local events like \textit{`Cardiff Film \& Comic Con'} or larger, international events like \textit{`Gamescom'}. In recent years, events like these have become increasingly popular; both drawing more attendees and spreading internationally.  San Diego Comic-Con is one of the largest  conventions in the world, with an attendance of over 130,000. This convention has consistently seen a 15\% growth for the last 17 years, resulting in about \$19 million in revenue as of 2017  \cite{DanielleChiriguayo2018}.

 
 \begin{figure}[b!]
	\centering
	\includegraphics[scale=0.36]{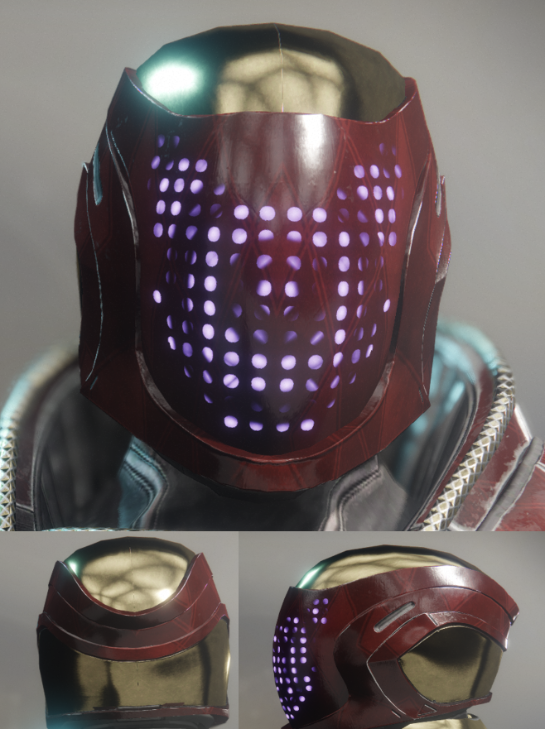}
	\caption{A Skullfort: This a popular iconic helmet in the pop culture. It is only appear in video games. }
	\label{Front1}
\end{figure}

Cosplay is a visual art form that is usually displayed in very public, crowded venues. As such, depending on their skill; these artists tend to receive some recognition and fame within the community. Since the coining of the term `Cosplay' in 1984 \cite{Bruno2002}, this community has experienced exponential international growth. Due to the growth of this sub-culture, many of these individuals have successfully converted the hobby into a professional business model by capitalising on their social media recognition and influence. Usually through advertisement, and voluntary support like Patreon or Koffi. 

Considering the rapid expansion of both industries (Internet of Things (IoT) \cite{Perera2015a} and Cosplay) and the tendency of creators to adopt new technologies and techniques, it is inevitable that IoT technology would slowly be incorporated into the cosplay community. This incorporation is still in its infancy, only being utilised by the most dedicated and ambitious costume creators. Currently, most of this technology is limited to basic functionality, such as LED lighting or basic animatronics. Most of this functionality is achieved by utilising some of the more basic wearable technology. Furthermore, in most of these scenarios, typical interaction pattern is that the costume responds to its environment or the player's behaviour (e.g., the colour of the lights may get changed when the player moves hands). In contrast, our research focuses on exploring scenarios where the audience (third party) get to manipulate the costume behaviour (e.g., the audience get to change the colour of the Skullfort using a mobile application). The possibilities of audience-influenced cosplays are endless.  However, as technology continues to improve and become more widely adopted, new issues and concerns must be considered. 

These concerns are mainly due to the public nature of cosplay and how this technology may interact and impact the general public. Such concerns could include negatively affecting the disabled, creating distractions, violence, hate speech, etc. As we discuss later in this paper, each of these possibilities brings new challenges and require protection mechanisms.


\section{IoT Skullfort: System Prototype}


As shown in Figure \ref{Front1}, we selected a popular iconic helmet in the pop culture called \textit{`Skullfort'}. By design, it allows us to add more meaningful audience influenced interactions using colour LEDs and LED-based text. We decided that the LED-based display would be the best way to provide enhanced capabilities to the audience.

\begin{figure}[h!]
	\centering
	\includegraphics[scale=0.085]{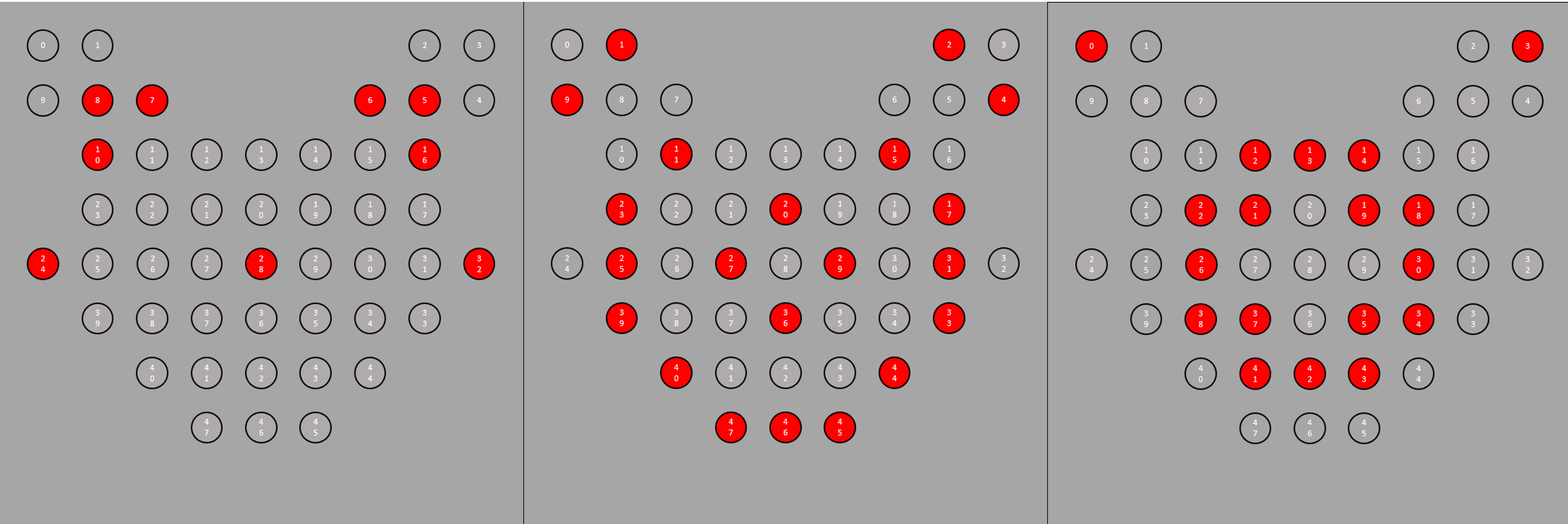}
	\caption{LED Matrix}
	\label{Matrix}
\end{figure}

As shown in Figure \ref{Matrix}, we developed a matrix using 60 LED per meter strip. These LEDs were individually addressable. Arduino-compatible Adafruit Feather nRF52 Bluefruit LE-nRF52832 was chosen as the micro-controller to manage the Bluetooth Low-Energy based communication. This chip is lightweight and portable, measuring only 51mm x 23mm x 8mm, which is adequately small enough to be housed within the helmet. Despite its small size, this chip is capable of being utilised as both a main microcontroller and a Bluetooth Low-Energy interface, meaning it can communicate with BLE devices, process the incoming data and control any connected peripherals. However, due to the smaller size of the device, it can only output 3.3V; which is insufficient for this project. As such, the `PowerBoost 1000'  was required. This device is a DC/DC boost converter module that can be powered by any 3.7V Lithium Polymer battery, and convert the battery output to 5.2V.

While lopping through an LED animation, this device continually listens for any new communication.  Once the device understands which command it has received, it can translate the given data into RGB colour values; ready to be utilised in the next animation stage. The microprocessor will then multiply the colour values by a strength value that is determined by the current elapsed time.



\begin{figure}[h!]
	\centering
	\includegraphics[scale=0.23]{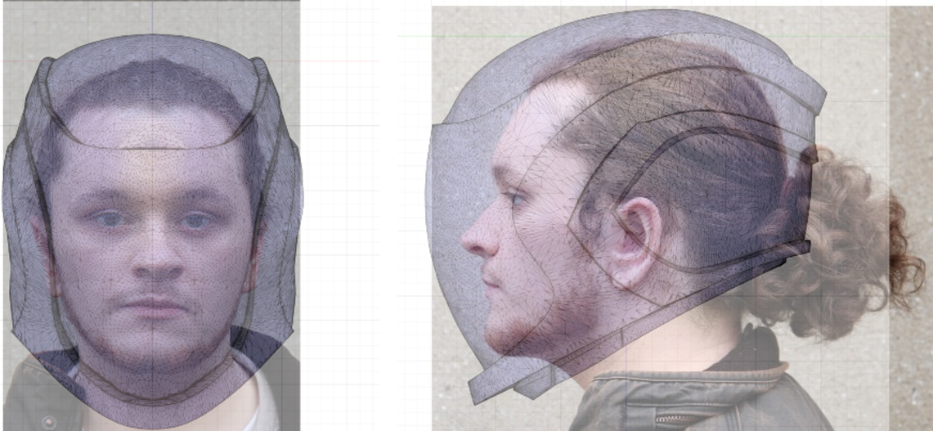}
	\caption{Scaled Reference Images}
	\label{Person}
\end{figure}

\begin{figure}[h!]
	\centering
	\includegraphics[scale=0.17]{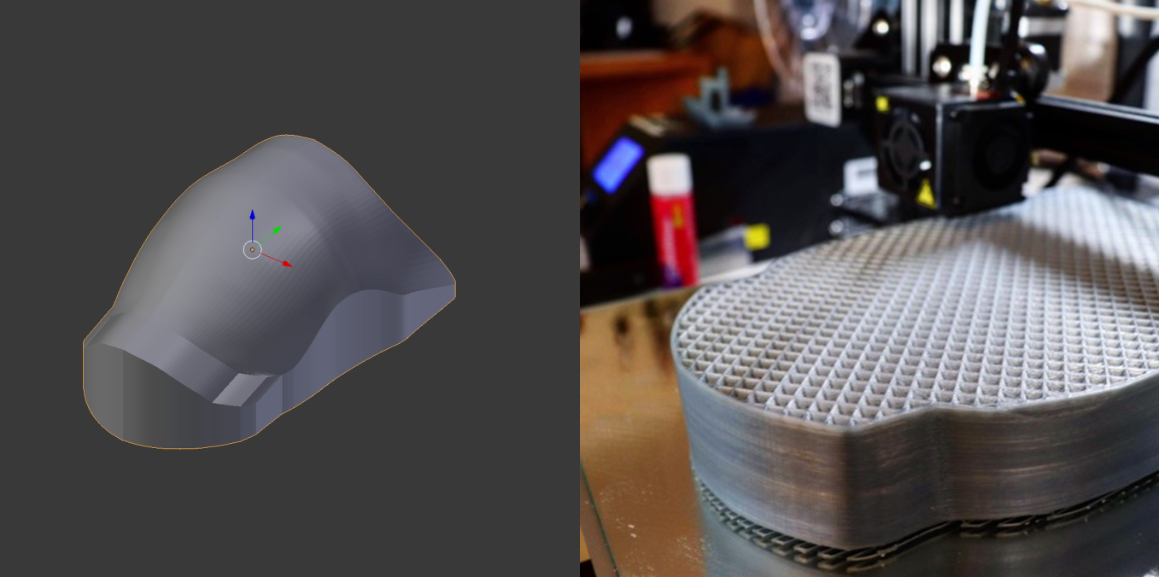}
	\caption{3D Printing Visor Mould}
	\label{3Dmodel2}
\end{figure}

\begin{figure}
	\centering
	\includegraphics[scale=0.42]{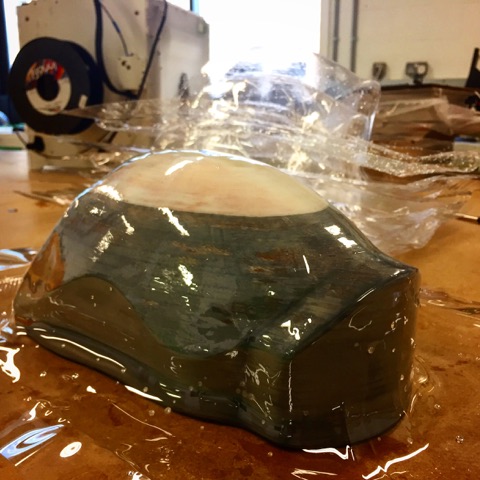}
	\caption{Vacuum Formed Visor}
	\label{VacuumFormedVisor}
\end{figure}

\begin{figure}
	\centering
	\includegraphics[scale=0.30]{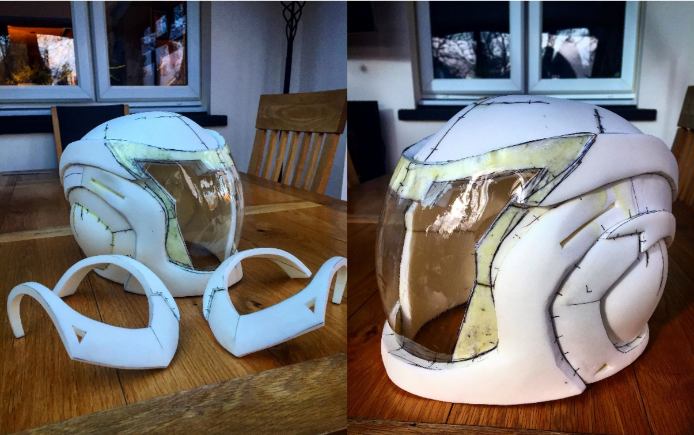}
	\caption{Fabricated Helmet}
	\label{FabricatedHelmet}
\end{figure}

\begin{figure}
	\centering
	\includegraphics[scale=0.44]{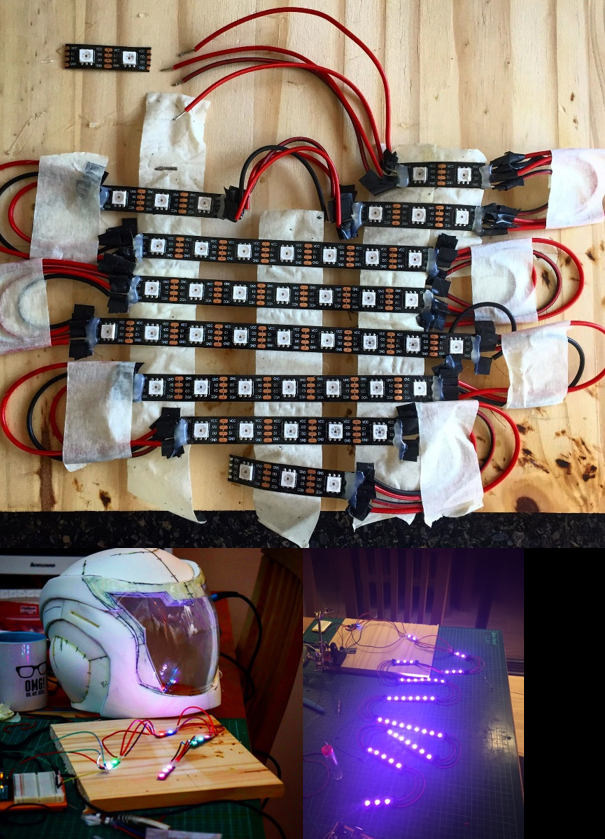}
	\caption{Completed LED Circuit}
	\label{CompletedLEDCircuit}
\end{figure}

\begin{figure}[t!]
	\centering
	\includegraphics[scale=0.34]{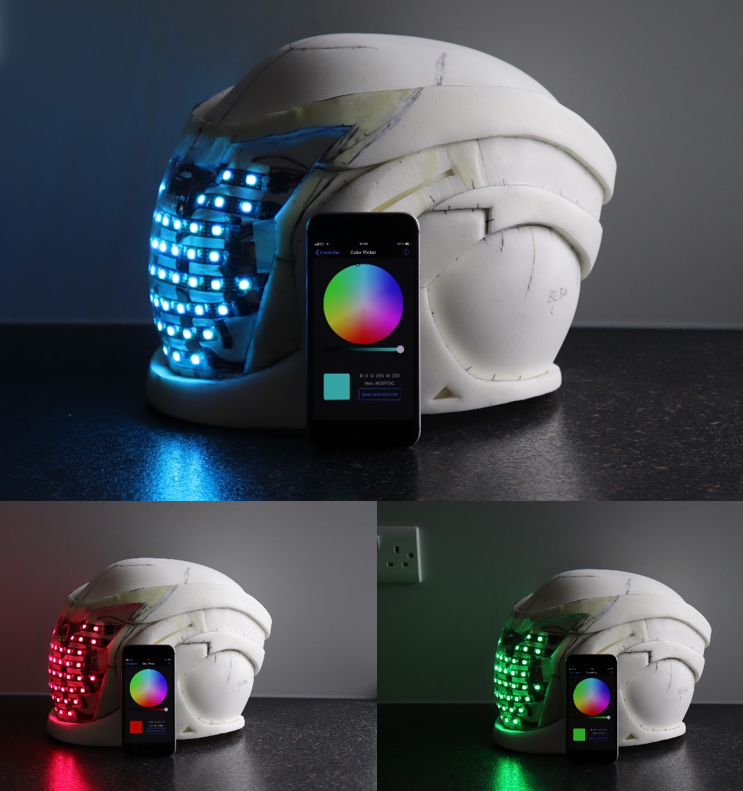}
	\caption{Completed Circuit and Helmet}
	\label{FinalHelmet}
\end{figure}

Blender which relies on vertices, edges and polygons to produce a 3D mesh was used to 3D model the prototype. As this method of modelling allows the transformation of individual vertices, it is far more suitable for modelling abstract/organic models.  Once the model was complete, it could be separated into individual parts for further processing. Firstly, a transparent visor was required to allow the light of the LEDs to shine through the helmet. A sheet of PETG plastic was vacuum formed to produce the the visor. Secondly, It was decided that the helmet would be fabricated from EVA foam. \textit{`Ethylene-vinyl Acetate'} is a dense, foam-like material that is commonly utilised in the cosplay community. A software called `Pepakura' which automatically flattens a 3D object into 2D templates has been used to fabricating the helmet. Series of Figures, namely, \ref{Person}, \ref{3Dmodel2}, \ref{VacuumFormedVisor}, \ref{FabricatedHelmet}, \ref{CompletedLEDCircuit}, illustrates different steps of the development process. The finished artifact is depicted in Figure \ref{FinalHelmet}.

\section{Focus Group Discussion: Opportunities and Challenges}

By using the IoT Skullfort prototype as a concrete example, we conducted a focus group to identify potential impact. Six experts with decades of experience from the cosplay community were invited. The focus group began with a short demonstration of the IoT Skullfort.  We invited experts to play both player and audience roles. We extracted the following themes during the discussion.

\subsection{{Learning curve for the audience}} If this technology were to be utilised in cosplay event, such a costume would enhance the performance. However, it could be challenging to educate the audience on how to interact with IoT costumes as there aren't any well-established interaction patterns.

\subsection{{Drawing attention in public events}}  If the cosplayer wearing the device seeks attention, then it is a huge advantage, capturing the focus of most around. However, if they do not seek attention, then the drawn crowd would be an irritation. Comparably, such a device can also impact the experiences of those in the immediate vicinity, both in a positive or negative manner. If the attendees are interested in such costumes, then it's a pleasure to view and interact with. However, there will be individuals with no interest. They will instead be hindered by the crowd that forms.

\subsection{{Increase of complexity and handling groups}}   The utilisation of IoT technology allows for another layer of creativity to be applied.  An interesting challenge would be to develop techniques to facilitate larger audiences to collectively engage with IoT costumes (i.e., voting system). Such technique should be correctly implemented with fair treatment for all the audience members. Otherwise, it could would harm the reputation of the cosplayer.



\subsection{{Health and Safety}} It was highlighted, that if tailored, this technology could be utilised to mitigate specific disabilities. For example, an individual who suffers from partial deafness could utilise a series of microphones scattered across a costume. However, there could be unfortunate accidents involving the unintentional interaction between the device and a disabled individual. For example, the flickering of the LEDs on this helmet could if erratic enough, possibly trigger an epileptic fit. Fortunately, one of the main strengths of these devices is the responsive interactivity; if such an individual expressed their discomfort, the device state could be quickly altered to accommodate their needs.

\subsection{{Inappropriate, abusive comments, hate speech}}
If the LEDs were utilised as a matrix that allowed scrolling text, then allowing open access to the public could cause trouble. Experts agreed that the individuals wearing the device is just as liable for the content displayed, as those that wrote it. Meaning, they would be punished for any problematic content; as they facilitated the public distribution of the message. 

Their opinions are supported by the event that occurred at Seattle University in  2018 where a YouTuber live-streamed video, utilised a text-to-speech function that allowed individuals watching to broadcast a message of their choosing from a speaker hidden within his jacket. One such viewer decided to broadcast a fake bomb alert and subsequent countdown to create panic \cite{Pineros2018}. To ensure that no such situation can occur again, the group recommended limiting the choices available to the audience. This would most likely either involve a whitelist of words, a collection of pre-approved sentences and states or utilising an individual to screen and approve messages before they are displayed. 
Even before the legality of the issue can be considered, the experts considered and agreed that this system could only display messages and images that conformed to the events guidelines.  Considering these are public events, held by private companies; they reserve the right to remove anyone from the premises. As such, care should be taken to not cause offense.


%

%


\section{Conclusion} 

Cosplay community needs to develop a strategy to address challenges in IoT costumes using techniques such as design principles, community guidelines, and technical means. More specifically, we expected that these guidelines to be co-developed in collaboration with domain experts (health and safety, legal, law enforcement, media and entertainment).

\begin{acks}
Charith Perera's work is supported by EPSRC PETRAS 2 (EP/S035362/1)
\end{acks}

\bibliographystyle{ACM-Reference-Format}
\bibliography{library}


\end{document}